\documentclass[twocolumn,prl,superscriptaddress]{revtex4-1}
\usepackage{palatino}
\usepackage[latin1]{inputenc}
\usepackage{epsf}
\usepackage{amsmath,amssymb}
\usepackage{latexsym}
\usepackage{calc}
\usepackage{color}
\usepackage{shadow}
\usepackage{epsfig}
\usepackage[usenames,dvipsnames]{xcolor}

\def\bea{\begin{eqnarray}}
\def\eea{\end{eqnarray}}
\def\ben{\begin{equation}}
\def\een{\end{equation}}
\def\benu{\begin{enumerate}}
\def\enu{\end{enumerate}}

\def\dulR{{\underline{\underline{\bf R}}}}
\def\dulr{{\underline{\underline{\bf r}}}}

\def\1var{(\bx_1...\bx\N)}

\def\br{{\bf r}}

\def\bx{{\br t}}

\begin{document}
\title{The Exact Potential Driving the Electron Dynamics in Enhanced Ionization}
\author{Elham Khosravi}
\email{elham.etn@gmail.com}
\thanks{Corresponding author} 
\author{Ali Abedi}
\email{aliabedik@gmail.com}
\address{Department of Physics and Astronomy, Hunter College and the Graduate Center of the City University of New York, 695 Park Avenue, New York, New York 10065, USA}
\address{Nano-Bio Spectroscopy Group and European Theoretical Spectroscopy Facility (ETSF), Universidad del Pa\'is Vasco CFM CSIC-UPV/EHU-MPC and DIPC, Av. Tolosa 72, 20018 San Sebasti\'an, Spain}
\author{Neepa T. Maitra}
\email{nmaitra@hunter.cuny.edu}
\address{Department of Physics and Astronomy, Hunter College and the Graduate Center of the City University of New York, 695 Park Avenue, New York, New York 10065, USA}

\date{\today}

\begin{abstract} 
It was recently shown that the exact factorization of the electron-nuclear wavefunction allows the construction of a Schr\"odinger 
equation for the electronic system, in which the potential contains exactly the effect of coupling to the nuclear degrees of freedom and any
external fields. Here we study the exact potential acting on the electron in charge-resonance enhanced ionization in a model
one-dimensional H$_2^+$ molecule. We show there can be significant
differences between the exact potential and that used in the
traditional quasistatic analyses, arising from
non-adiabatic coupling to the nuclear system, and that these are crucial to
include for accurate simulations of time-resolved
ionization dynamics and predictions of the ionization yield. 
\end{abstract}
\maketitle 
Ionization is a fundamental process in the strong-field
physics of atoms and molecules, lying at the heart of many
fascinating phenomena such as high harmonic generation, Coulomb
explosion, laser induced electron diffraction, and molecular orbital
tomography. The ionization rate from a molecule can be several orders
of magnitude higher than the rate from the constituent atoms at a critical range of internuclear separations. 
This phenomenon, termed charge-resonance enhanced ionization (CREI), was
first theoretically predicted about twenty years ago~\cite{ZCB93,ZB95,SIC95,CCZB96}, and soon after verified experimentally~\cite{BWSC08,CSC96,WMS12}.

The enhancement in the ionization rate has been explained by a
quasistatic  argument in the pioneering works of
Ref~\cite{ZB95,CFB98,CB95,CCZB96,YZB95,BL12}, treating the
nuclei as instantaneously-fixed point particles, with the electrons following the combined potential from the laser field and the electrostatic attraction of the nuclei. 
In any experiment however, the nuclei are neither frozen, nor are they
point particles;  instead their motion can be {\it strongly} coupled to the electron
dynamics and accounting for the coupled electron-ion quantum dynamics can be essential~\cite{HS08}. Further, the electron does not simply follow the field
adiabatically, evidenced by the multiple subcycle ionization bursts recently revealed in Refs.~\cite{TB10,TB11}. 
A handful of calculations treating the full dynamics of
quantum nuclei and the electron in H$_2^+$~\cite{CCZB96,CFB98}, as well as a few experiments, have
verified that  the essential CREI phenomenon remains
robust, although the details are altered by the ionic dynamics.  For
example, in Refs.~\cite{BWSC08} it was found that nuclear
motion washes out the two-peak structure predicted in the
frozen-nuclei analysis of CREI~\cite{ZB95} into a single broad peak.
Further,  the CREI enhancement is subdued if, during the
experiment, only little of the nuclear density reaches the critical
internuclear separation~\cite{LLDQ03}. Hence, to properly  understand, model, and predict the experiment, a fully
time-dependent picture of coupled electronic and ionic motion is
needed. 

In this Letter, we utilize the exact factorization approach~\cite{AMG10,AMG12,SAMYG14,SAMG15} to investigate the
dynamics of the electron during the CREI process, as influenced by the
field and nuclear wavepacket dynamics. In particular, we study the exact time-dependent potential that drives the electron and
fully accounts for coupling to both the field and the dynamical
nuclei. This concept has been introduced by the exact factorization
in its inverse form~\cite{SAMYG14}, and here we show that, for the
CREI process, this exact potential can be remarkably different from 
the quasistatic potential, or even from a modified quasistatic potential that accounts for the width and splitting of the nuclear
wavepacket. This indicates the need for dynamical electron-nuclear correlation effects to be included in the calculation.  
Further, we identify a new measure of ionization appropriate for fully
dynamical studies, which indicates the regions of the nuclear wavepacket associated with the ionizing electron.

By
restricting the motion of the nuclei and 
the electron in the $H_2^+$ molecule to the polarization direction  of the laser field 
 the problem can be modeled with a one-dimensional Hamiltonian featuring ``soft-Coulomb'' interactions~\cite{JES88} (atomic units, $e = m = h= 1$, are used throughout
the article, unless otherwise noted):
\begin{eqnarray}
  \label{eq:H2+}
    \hat{H}(t) &=&  - \frac{1}{2\mu_e}\frac{\partial^2}{\partial z^2} - \frac{1}{M}\frac{\partial^2}{\partial R^2}- \frac{1}{\sqrt{1+(z-R/2)^2}} \nonumber \\
       &-&  \frac{1}{\sqrt{1+(z+R/2)^2}} + \frac{1}{\sqrt{0.03+R^2}} + \hat{V}_l(z,t)      
\end{eqnarray}
where $R$ and $z$ are the internuclear distance and the electronic
coordinate as measured from the nuclear center-of-mass, respectively.
The proton mass is denoted as $M$ while $\mu_e=(2M)/(2M+1)$  is the electronic reduced mass. 
The laser field, within the dipole approximation, is represented by
$\hat{V}_l(z,t) = q_e \,z\,E(t)$ where $E(t)$ denotes the electric field
amplitude and the reduced charge $q_e =(2M+2)/(2M+1)$.
Such reduced-dimensional models have been shown to qualitatively reproduce experimental results (see Ref.~\cite{KMS96} for example).
Here we first study the dynamics of the system subject to a $50$-cycle
pulse of wavelength $\lambda\,= 800$ nm ($\omega=0.0569$ a.u.)  and
intensity $I = 2 \times 10^{14}$W/cm$^2$, with a sine-squared pulse
envelope.

\begin{figure}
\begin{center}
  \hspace*{-0.35 cm}\includegraphics[width=0.845\columnwidth]{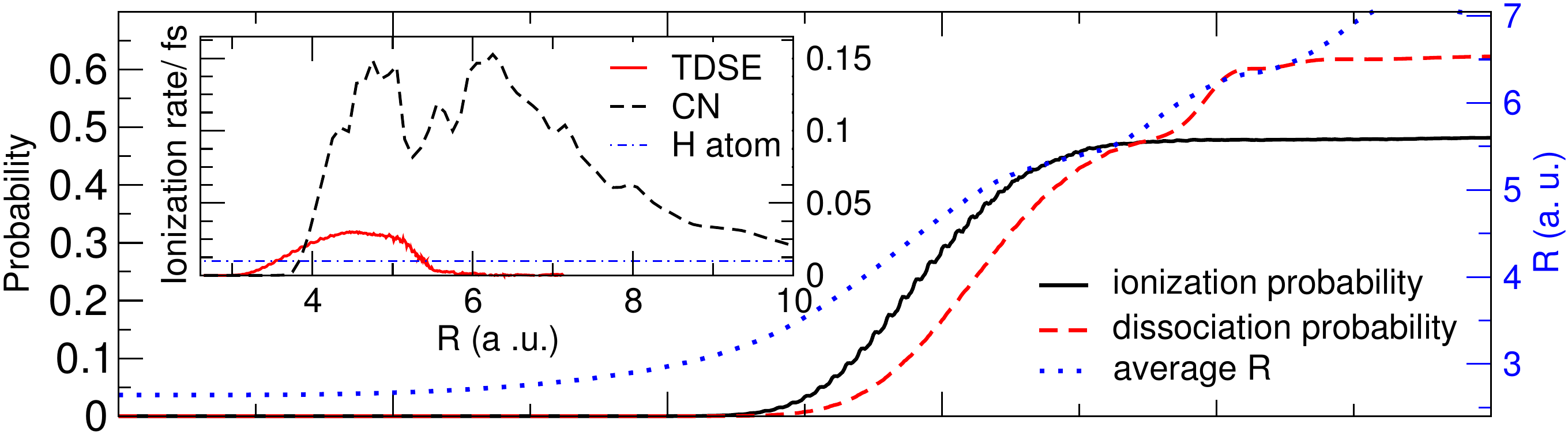}
\includegraphics[width=0.95\columnwidth]{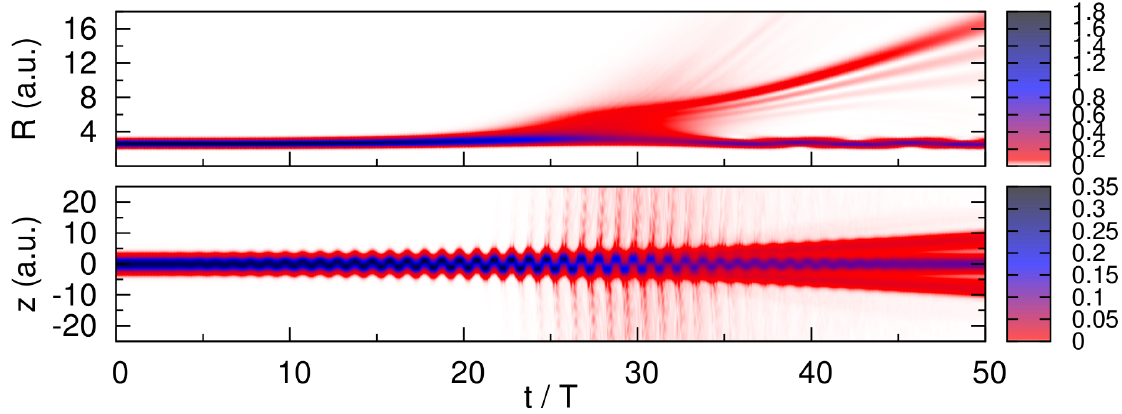}
\end{center}
\caption{Upper panel: Ionization probability , dissociation probability , average internuclear distance ,$\langle R \rangle $, as a function of number of cycles $t/T$.
The inset depicts the ionization rate for: clamped-nucleus (CN) calculation, as a function of the internuclear distance (black dashed curve),
the exact molecular TDSE as a function of $\langle R \rangle $ (red full curve), and the H atom (dash dotted line). Middle panel:
contour plot of the time-dependent nuclear density.  Lower panel: contour plot of time-dependent electronic density.}
\label{fig:fig1}
\end{figure}

Setting the ground-state of the molecule as the initial
state, we first solve the time-dependent Schr\"odinger
equation (TDSE) for Hamiltonian in Eq. (\ref{eq:H2+}) numerically exactly. 
The upper panel of Figure~\ref{fig:fig1},  shows the ionization probability, dissociation probability and the average internuclear 
distance $\langle R \rangle $ as a function of number of optical cycles ($t/T$)~\footnote{ These probabilities are calculated as defined in~\cite{CCZB96}. 
Here we have chosen $z_I = \pm 15$ for ionization box while the dissociation box is defined as $R_D < 10 $. 
}. Here $T$ denotes  duration of one cycle which in this case is $T=2.67$ fs.
We see from the figure that ionization is rapidly onset as we approach
the middle of the pulse, slowing down later while the field
decreases. The nuclei dissociate primarily via Coulomb explosion
following ionization. Note that the majority of the ionization occurs
when $\langle R \rangle$ is between 4--5.5 a.u. The inset of the upper panel of Figure (\ref{fig:fig1}) shows the
ionization rate which in the fully dynamical calculation, is plotted against the  average internuclear separation $\langle R \rangle (t)$, while in the clamped-nuclei (CN) case,
 it is calculated and plotted for each fixed internuclear separation $R$
~\footnote{for the  fully dynamical calculation (labelled TDSE in legend) the ionization rate at time $t$ is calculated via $ (\ln( p_v(t-T/2)) -\ln(p_v (t+T/2)))/T$ and 
plotted against the corresponding average internuclear separation $<R>(t)$ at time $t$. Here $p_v$ is the probability inside the ionization box.  In case of CN the 
ionization rate is computed as given in \cite{CCZB96}.}.
In the CN case, the peak near ~6.5 a.u. is usually identified with CREI 
while that near ~5 a.u. is associated with a symmetry-breaking electron localization~\cite{ZB95}. 
The exact ionization rate, however, has a single broad peak centered between 4 a.u. and 5 a.u., and is smaller than that of the CN calculations, but still higher than the atomic 
rate, similar to the observations in Refs.~\cite{CCZB96, BWSC08}.

One should keep in mind that representing the ionization rate in the fully-dynamical case as a
function of just the average internuclear
separation has however only limited meaning: the nuclear charge
distribution, plotted in the middle panel of
Figure~\ref{fig:fig1}, bifurcates, so considering ionization simply as a function
of the average separation does not properly indicate the
internuclear separations at which the ionization rate is enhanced.
A large fragment of the nuclear density remains localized, oscillating
around the equilibrium separation, while another part begins to dissociate, soon after the ionization is onset. 
This can be seen clearly in the  plot of the electronic density  plotted in the lower panel of
Figure~\ref{fig:fig1}.
Hence a  dynamical picture of the CREI that accounts for the coupling to the 
nuclear distribution as it changes in time is desirable. 

A complete picture of the electronic dynamics coupled to the non-classical nuclei is provided within the exact
factorization framework in its {\it inverse} formulation~\cite{SAMYG14}:
Complementary to the direct factorization of Refs.~\cite{AMG10,AMG12},
the exact electron-nuclear wavefunction $\Psi(\dulr,\dulR,t)$ that solves
 the full electron-nuclear TDSE can be exactly written as a
product $\Psi(\dulr,\dulR,t)=\Phi(\dulr,t)\chi_\dulr(\dulR,t)$, where
$\Phi(\dulr,t)$ may be interpreted as the electronic wavefunction and
$\chi_\dulr(\dulR,t)$ the conditional nuclear wavefunction that
parametrically depends on the electronic configuration $\dulr$ and
satisfies the partial normalization condition $\int d\dulR
\vert\chi_\dulr(\dulR,t)\vert^2 =1$ for every $\dulr$ at each $t$. The
electronic wavefunction can be shown to yield the exact $N_e-$body
electronic density, and $N_e$-body electronic current-density of the
system.  The equations of motion that the electronic and nuclear
factors satisfy are presented in~\cite{SAMYG14}. The electronic
equation, in particular, has the appealing form of a TDSE
 that contains an exact time-dependent scalar potential as
well as a time-dependent vector potential: for any
one-dimensional case, we can choose a gauge such that the vector potential is zero~\cite{SAMYG14,AMG10,AMG12}, and then:
\ben
\label{eq:eleq-if}
\left(-\frac{1}{2\mu}\frac{\partial^2}{\partial z^2}+\epsilon_e(z,t)\right)\Phi(z,t)=i\partial_t\Phi(z,t).
\een
Hence, the exact potential driving the electron dynamics is $\epsilon_e(z,t)$ which can be compared with the
traditional potentials used to study electronic dynamics. This potential, termed the time-dependent
potential energy surface for electrons, $e$-TDPES, can be decomposed to 
\ben
\label{eq:epes_tot}
\epsilon_e(z,t) = \epsilon^{\rm app}(z,t) + \mathcal{T}_n (z,t)+ \mathcal{K}^{\rm cond}_e(z,t) +  \epsilon^{\rm gd}_e(z,t),
\een 
where for our H$_2^+$ model,
\bea
\label{eq:eps_approx}
\epsilon^{\rm app}(z,t) &=& \left\langle\chi_z(R,t) \right\vert 
 \hat{W}_{en}(z,R)+\hat{W}_{nn}(R) \left\vert
 \chi_z(R,t)\right\rangle_R  \nonumber \\
 &+&\hat{V}^{l}(z,t) ,
\eea
is an approximate potential that generalizes the traditional quasistatic potential to the case of a quantum nuclear wavepacket. The second term,
\ben
\mathcal{T}_n(z,t) = -\left\langle\chi_{z}(R,t) \right\vert \partial_R^2\left\vert \chi_{z}(R,t)\right\rangle_R/M\;,
\label{eq:eps_Tn}
\een
represents a nuclear-kinetic contribution to the electronic potential from the
conditional nuclear wavefunction, while,
\ben
\mathcal{K}^{\rm cond}_e(z,t)= \langle \partial_z\chi_{z}(R,t) \vert \partial_z \chi_{z}(R,t)\rangle_R/M\;,
\label{eq:eps_Ke}
\een
represents an electronic-kinetic-like contribution from the conditional nuclear wavefunction, and,
\ben
\epsilon^{\rm gd}_e(z,t)= \left\langle\chi_{z}(R,t)\right\vert - i \partial_t\left\vert \chi_{z}(R,t)\right\rangle_R\;,
\label{eq:eps_gd}
\een
is the gauge-dependent component of the potential. It is important to note that $\epsilon^{\rm app}$ reduces to the quasistatic potential
 when the nuclear density is approximated classically as a $z$-independent delta-function at $\bar{R}(t)=\langle R \rangle (t)$:
\ben
\label{trad_tdpes}
\epsilon^{\rm qs}(z,t|\bar{R}(t))= \hat{W}_{en}(z,\bar{R}(t)) +
\hat{W}_{nn}(\bar{R}(t))+ \hat{V}^{l}(z ,t).  
\een 

\begin{figure}
\begin{center}
\includegraphics[width=8.50cm]{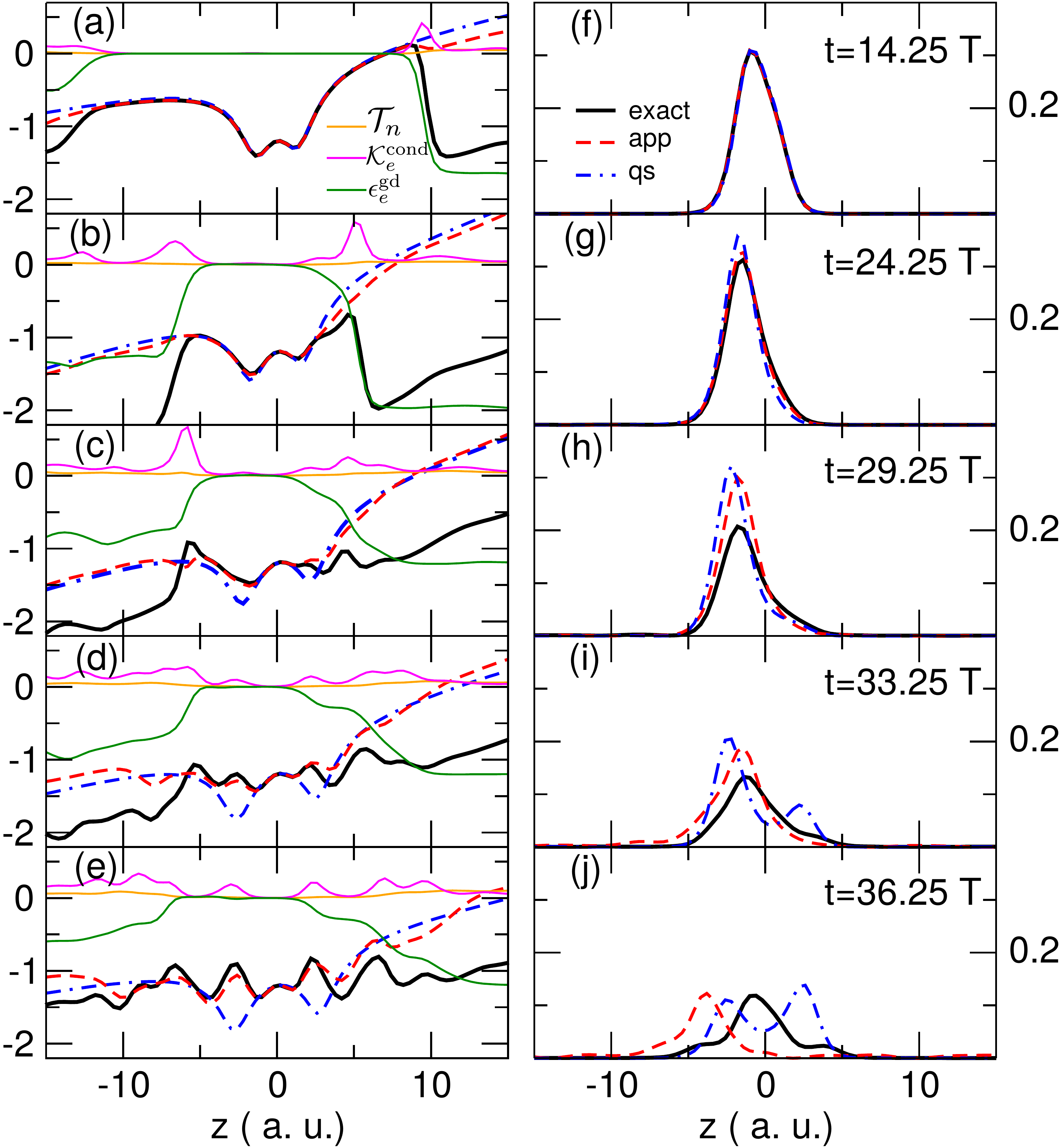}
\end{center}
\caption{The exact electronic potential $\epsilon_e$ (black solid line) and its various components together with 
the quasistatic potential $\epsilon^{\rm qs}$ (blue dash line) are plotted in the left-hand side at five different snapshots of time. The exact 
electron density together with the electron density calculated from propagating the electron on $\epsilon^{\rm qs}$ and $\epsilon^{\rm app}$
are plotted on the right-hand side. }
\label{fig:fig2}
\end{figure}

\begin{figure}
\begin{center}
\includegraphics[width=1\columnwidth]{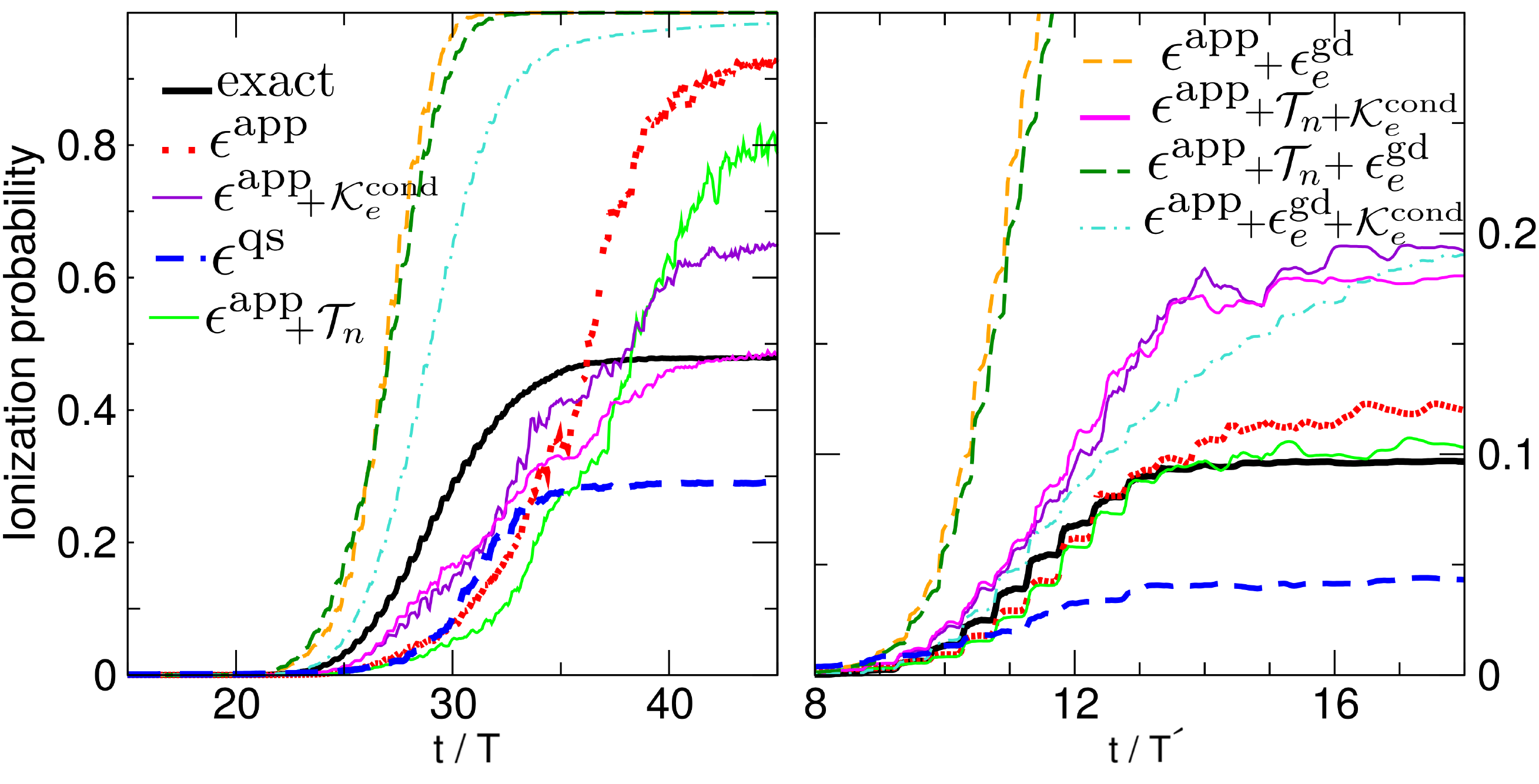}
\end{center}
\caption{Ionization yields calculated from propagating the electron on different components of
the exact electronic potential as well as on the quasistatic potential. Left: $\lambda\,= 800$ nm and intensity $I = 2
  \times 10^{14}$ W/cm$^2$ ($50$-cycle). Right:  $\lambda\,= 600$ nm and intensity $I = 10^{14}$W/cm$^2$
  ($20$-cycle). Legends apply to both.}
\label{fig:fig3}
\end{figure}

 We now investigate the $e$-TDPES~(Eq.~\ref{eq:epes_tot}) and discuss
 the impact of its
 components~(Eqs.~\ref{eq:eps_approx}--\ref{eq:eps_gd}) on the
 electron dynamics. In particular, we ask how well electron
 propagation on $\epsilon^{\rm app}$ performs: is accounting for the
 width of the nuclear wavepacket, and its correlation with the
 electron dynamics via the parametric dependence, enough to capture
 accurately the full electron dynamics?  In Figure~\ref{fig:fig2} the
 exact $e$-TDPES, $\epsilon_e$, (black solid line) and its
 various components together with the quasistatic potential $\epsilon^{\rm
   qs}$ (blue dash line) are plotted on the left-hand side at five
 different snapshots of time  in which the field is at the maximum of the cycle. The exact electron density together with
 the electron density calculated from propagating the electron on
 $\epsilon^{\rm qs}$ and $\epsilon^{\rm app}$ are plotted on the
 right-hand side. 
In all these calculations, the initial state is chosen to be the
 exact ground state density of the complete system. 
We also plot the ionization yields calculated from propagating the electron on different components of
the exact electronic potential on the left on Fig.~\ref{fig:fig3}.

We have chosen times that are representative of three different phases
of the dynamics (refer to Fig.~\ref{fig:fig1}): (1) the initial phase, up to $t\approx20\,T$, for which the
dissociation and ionization probabilities are still negligible (panel (a) of Fig.~\ref{fig:fig2}, shown at $t = 14.25\,T$), (2) the second phase,
$\sim 20\,T< t < \sim 35\,T$,
which is when
ionizaton/dissociation mostly occurs; panels (b)--(d) of Fig.~\ref{fig:fig2} ($t= 24.25, 29.25, 33.25\,T$) 
are chosen to illustrate this phase, (3) the final phase, $t> ~35\,T$, in which the system begins to stabilize, while
the field intensity decreases, represented at  $t = 36.25\,T$ in panel (e).

In the first phase of the dynamics (representative panel (a)), the nuclear wavepacket is quite
localized around its initial position and the $e$-TDPES, quasistatic
potential and approximate potential are essentially on top of each other in the central
region ($|z|<10$ a.u.). They differ from each other only in the tail of the electronic density,
where, in particular, $\epsilon_e$ has a large step downward
(Fig.~\ref{fig:fig2}(a)). Since the density is very tiny in the tail 
region, the overall dynamics is not affected significantly by this
feature. In the second phase of the dynamics 
the nuclear motion begins to pick up, affecting the shape of the exact
$e$-TDPES in the central region.  From this point on the exact
potential begins to develop features that are absent in the
quasistatic potential. As part of the nuclear density begins to
stretch apart, the $e$-TDPES begins to exhibit a double well structure
in the up-field side 
of the potential ($0<z<5$ a.u.), while the down-field side maintains a
single well structure as is shown in panels (b) and (c). Further, the
depth of the central wells are decreased compared to the quasistatic
picture. Outside the central region ($|z|>5$ a.u.) the $e$-TDPES drops
down, yielding a barrier that is smaller and narrower than that of the
quasistatic potential. This feature in particular, significantly
facilitates the tunnelling ionization of the electron density in the
exact dynamics already at $t= 24.25\,T$, evident in the spreading of the exact density (panel (g), see also left
panel of Fig.~\ref{fig:fig3}).  In the case of the quasistatic and
approximate potentials, the ionization is still
negligible at this time, due to small tunnelling probability.
The differences between the exact $e$-TDPES and both $\epsilon^{\rm app}$ and $\epsilon^{\rm qs}$ continue to grow in the central region ($|z|<5$ a.u.) 
throughout the second phase (panels (b)--(d), and corresponding electronic densities (g)--(i)),  as contributions from $\epsilon_e^{\rm gd}$  and
$\mathcal{K}_e^{\rm cond}$ get larger and extend closer to the center.
 It is interesting to note that $\epsilon_e^{\rm gd}$ typically has large
steps that lowers the potential on both sides, allowing for more
ionization (see also Fig.~\ref{fig:fig3}, left panel), while $\mathcal{K}^{\rm cond}_e$ develops several (smaller) barrier structures, 
whose net effect also appears to increase the ionization yield in this phase (see Fig.~\ref{fig:fig3}).
The $\mathcal{T}_n$ term has very small barriers in the outer region whose tendency is to confine the density, leading to a decrease in the ionization probability.

By the end of the second phase, at $t = 33.25\,T$ (Fig.~\ref{fig:fig2}
(d)), the exact potential is totally different from the quasistatic
potential, everywhere except for at small $|z|<~1$, presenting a shallow double well structure in both up-field
and down-field sides of the potential. Furthermore, the discrepancy
between the $\epsilon^{\rm qs}$ and $\epsilon^{\rm app}$ becomes more
noticeable as the nuclear wavepacket begins to split and dissociate
in the field. 
By this time, there has been significant ionization in all
three cases (left panel of Fig.~\ref{fig:fig3}), although more in the exact case, as discussed above.  
Towards the end of the second phase, the ionization yields of the quasistatic and approximate calculations begin 
to differ from each other, as expected from the growing discrepancy between their respective potentials.

Entering now the third phase of the dynamics, four optical cycles
later (panel (e) in Fig.~\ref{fig:fig2}), the exact potential differs dramatically from the other two
showing a formation of four wells in the central region ($|z|<6$
a.u.). The two wells in the center are associated with the part of the
nuclear density that is not dissociated while the other two are
associated with the dissociating fragment, and hence, they move
outwards. The $e$-TDPES consequently localizes the electronic density in
three positions as seen in panel (j) of Fig.~\ref{fig:fig2}, namely in the center
corresponding to the part of the nuclear density that is localized
around the equilibrium, and on each of the dissociating fragments of
protons. In the third phase, $\epsilon^{\rm app}$ grossly overionizes the system; as $\epsilon^{\rm app}$ 
has many shallow barriers and continues to oscillate in the field, failing to stabilize. The quasistatic 
potential $\epsilon^{\rm qs}$ retains a deep double well structure throughout the dynamics, in contrast to the exact; 
as the field reduces toward the end of the pulse the ionization in either of these cases saturates,  
but the quasistatic fails to get the density and ionization yield correct.

The left panel of Figure~\ref{fig:fig3} shows that neglecting all
the electron-nuclear correlation terms except for $\epsilon^{\rm app}$ underestimates the ionization at first, but later, as the exact
ionization begins to saturate, the ionization from $\epsilon^{\rm app}$ continues to grow, and leads
ultimately to a significant overestimate of the total ionization.  Even
propagating on the quasistatic potential, a crude approximation given
the earlier discussion, gives a better ionization yield.  We see from
Fig.~\ref{fig:fig3} (left) that adding $\mathcal{T}_n$ to $\epsilon^{\rm app}$ 
reduces the ionization probability at all times, due to its small confining barriers as mentioned above. On
the other hand, adding $\mathcal{K}^{\rm cond}$ to $\epsilon^{\rm app}$ increases the ionization at first, and then decreases
it, giving an overall somewhat improved prediction of the ionization dynamics
relative to dynamics on $\epsilon^{\rm app}$ alone. Although adding both $\mathcal{K}^{\rm cond}$  and $\mathcal{T}_n$ to $\epsilon^{\rm
  app}$ seems to give a good final ionization yield, the intermediate dynamics is not very good. Adding
$\epsilon_e^{\rm gd}$ to $\epsilon^{\rm app}$ drastically overshoots the ionization, yielding ultimately a completely ionized
molecule. It appears for the current choice of laser parameters and
initial state all these dynamical electron-nuclear correlation terms are important to include to obtain good prediction of the
ionization yield. But, is this conclusion general? 
Does $\epsilon^{\rm app}$ always perform so poorly? 

We performed the same calculations
using a $20$-cycle pulse of wavelength $\lambda\,= 600$ nm ($\omega=0.076$ a.u.)  and
intensity $I = 10^{14}$W/cm$^2$, with a sine-squared pulse
envelope. We further set the initial state to be the $6th$ excited
vibrational state (c.f.~\cite{CCZB96}). The resulting ionization yields computed from the
propagation of the electron using the different components of the exact
potential is presented in the right panel of Fig.~\ref{fig:fig3} ($T'=2$ fs, is the duration of one cycle.). We see that the result of electron dynamics on $\epsilon^{\rm app} $ in this case agrees very well with the
exact result, and further, that with the addition of $\mathcal{T}_n$ 
becomes even better. Other combinations of the potential components do not provide satisfactory
results. In particular, once again $\epsilon_e^{\rm gd}$ severely overestimates the ionization, while the the quasistatic dynamics underestimates the ionization
yield significantly. In this case,  the exact potential differs
substantially from the quasistatic potential from the very start, due to the vibrational excitation of the initial
state.  

\begin{figure}
\begin{center}
\includegraphics[width=1\columnwidth]{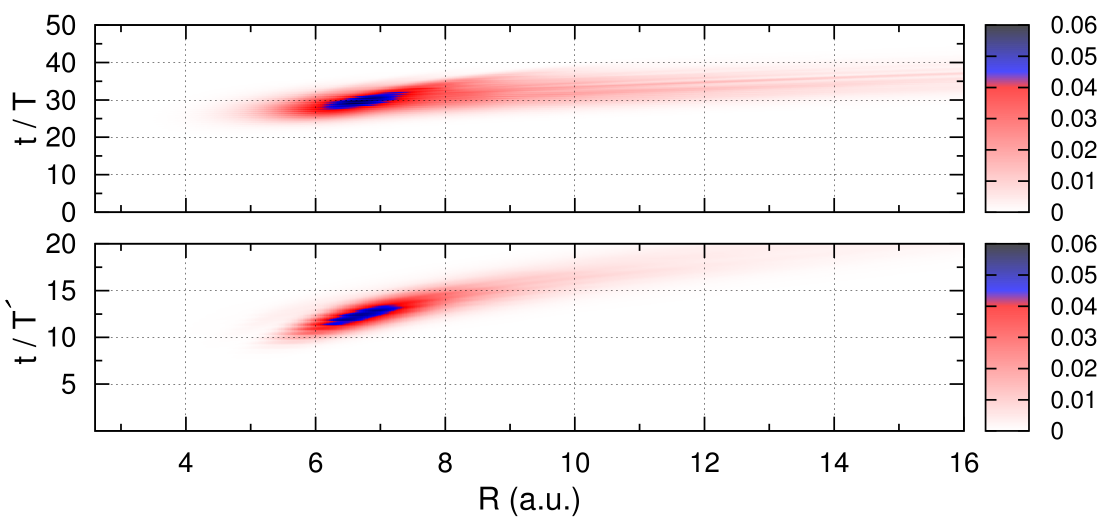}
\end{center}
\caption{Time-resolved, $R$-resolved, ionization probability, $I(R,t)$.
Upper panel:  $\lambda\,= 800$ nm and intensity $I = 2
  \times 10^{14}$ W/cm$^2$ ($50$-cycle). Lower panel: $\lambda\,= 600$ nm and intensity $I = 10^{14}$W/cm$^2$
  ($20$-cycle).}
\label{fig:fig4}
\end{figure}

From the discussion above it is clear that ionization dynamics depends
crucially on the coupling to the quantum nuclear motion; accounting
for both the splitting of the wavepacket as well as its dynamics is
important. At this point, one may ask from which part of the nuclear wavepacket is the ionization mostly occurring?
To answer this, we introduce a
time-resolved, $R$-resolved, 
ionization probability via
\ben
I(R,t) =\int_{z'_I}dz \vert\Psi(z,R,t)\vert^2\;,
\een
with  $\int_{z'_I} = \int_{-\infty}^{-z_{I}} +\int^{\infty}_{z_{I}}$ taking $z_{I} = 15$ a.\,u.,  
plotted in Fig~\ref{fig:fig4} for {\it both} of the laser parameters studied
in this work. In both cases, we observe a clear peak of $I(R,t)$, centered
around $6$ a.u.$ < R < 7.5$  a.u., the region predicted by the
quasistatic analysis of CREI, soon after the fields reach their maximum intensities . 
Hence, the quantity $I(R,t)$ 
represents a very useful measure of CREI in a fully dynamical picture, indicating clearly the dominant
internuclear separations at which ionization occurs. 
 We also point out that this quantity is analogous to the ionization probability at a given internuclear separation in
the quasistatic picture~\footnote{ $I(R,t)$ can be rewritten in terms of the exact conditional electronic density $|\Phi_R(z,t)|^2=|\Psi|^2/\int dz |\Psi|^2$ 
(see ~\cite{AMG10}), i.e, $I(R,t) =\int dz |\Psi|^2\left[1-\int_{z_{I}}dz|\Phi_R(z,t)|^{2}\right]$. Note that $|\Phi_R|^2$
gives the conditional probability of finding electrons at $z$ at time
$t$, given that the internuclear separation $R$ and can 
be viewed as an exactification of the electronic density for a given
internuclear separation in quasistatic picture. Therefore, $I(R,t)$ can be also
viewed as an exact concept analogous to the ionization probability in
the quasistatic picture.}. 

In summary, we have found the exact potential driving the electron dynamics in
a model one-dimensional H$_2^+$ molecule undergoing CREI. The presented potential  
provide a complete details of the CREI process beyond the quasistatic picture traditionally 
used to analyze and interpret this process. 
The large differences in the two potentials reveals the importance of dynamical electron-nuclear correlation terms lacking
in previous pictures of CREI. The drastic impact these can have on the
dynamics has been demonstrated, i.e. propagating the electrons in a
potential that neglects these terms lead to large errors in the
predictions of the ionization yield. Going beyond the quasistatic treatment by only accounting for the width and splitting of the nuclear wavepacket
  is generally not enough to get the correct dynamics of CREI.
How significant the dynamical electron-nuclear effects are for CREI phenomena in
larger systems~\cite{BKPB11} remains to be investigated, and moreover,
how to accurately model them in practical approximations will open a
major avenue for future research.
We further presented a time-resolved, $R$-resolved measure of CREI that takes into account the dynamical electron-nuclear correlation. This quantity
has a clear peak in the region predicted by the quasistatic analysis of CREI, for the fields studied here, soon after the fields reach their maximum intensities.

\begin{acknowledgments}
Financial support from the National Science Foundation
CHE-1152784 (N.T.M), and Department of Energy, Office of Basic Energy
Sciences, Division of Chemical Sciences, Geosciences and Biosciences
under Award DE-SC0008623  (E.K, A.A)  are gratefully acknowledged.
E.K and A.A also acknowledge financial support by the European Research 
Council Advanced Grant DYNamo (ERC- 2010-AdG-267374) and Grupo Consolidado UPV/EHU del Gobierno Vasco (IT578-13).
\end{acknowledgments}

\addcontentsline{toc}{section}{References}
\bibliography{./ref}

\begin{thebibliography}{25}%
\makeatletter
\providecommand \@ifxundefined [1]{%
 \@ifx{#1\undefined}
}%
\providecommand \@ifnum [1]{%
 \ifnum #1\expandafter \@firstoftwo
 \else \expandafter \@secondoftwo
 \fi
}%
\providecommand \@ifx [1]{%
 \ifx #1\expandafter \@firstoftwo
 \else \expandafter \@secondoftwo
 \fi
}%
\providecommand \natexlab [1]{#1}%
\providecommand \enquote  [1]{``#1''}%
\providecommand \bibnamefont  [1]{#1}%
\providecommand \bibfnamefont [1]{#1}%
\providecommand \citenamefont [1]{#1}%
\providecommand \href@noop [0]{\@secondoftwo}%
\providecommand \href [0]{\begingroup \@sanitize@url \@href}%
\providecommand \@href[1]{\@@startlink{#1}\@@href}%
\providecommand \@@href[1]{\endgroup#1\@@endlink}%
\providecommand \@sanitize@url [0]{\catcode `\\12\catcode `\$12\catcode
  `\&12\catcode `\#12\catcode `\^12\catcode `\_12\catcode `\%12\relax}%
\providecommand \@@startlink[1]{}%
\providecommand \@@endlink[0]{}%
\providecommand \url  [0]{\begingroup\@sanitize@url \@url }%
\providecommand \@url [1]{\endgroup\@href {#1}{\urlprefix }}%
\providecommand \urlprefix  [0]{URL }%
\providecommand \Eprint [0]{\href }%
\providecommand \doibase [0]{http://dx.doi.org/}%
\providecommand \selectlanguage [0]{\@gobble}%
\providecommand \bibinfo  [0]{\@secondoftwo}%
\providecommand \bibfield  [0]{\@secondoftwo}%
\providecommand \translation [1]{[#1]}%
\providecommand \BibitemOpen [0]{}%
\providecommand \bibitemStop [0]{}%
\providecommand \bibitemNoStop [0]{.\EOS\space}%
\providecommand \EOS [0]{\spacefactor3000\relax}%
\providecommand \BibitemShut  [1]{\csname bibitem#1\endcsname}%
\let\auto@bib@innerbib\@empty
\bibitem [{\citenamefont {Zuo}\ \emph {et~al.}(1993)\citenamefont {Zuo},
  \citenamefont {Chelkowski},\ and\ \citenamefont {Bandrauk}}]{ZCB93}%
  \BibitemOpen
  \bibfield  {author} {\bibinfo {author} {\bibfnamefont {T.}~\bibnamefont
  {Zuo}}, \bibinfo {author} {\bibfnamefont {S.}~\bibnamefont {Chelkowski}}, \
  and\ \bibinfo {author} {\bibfnamefont {A.~D.}\ \bibnamefont {Bandrauk}},\
  }\href {\doibase 10.1103/PhysRevA.48.3837} {\bibfield  {journal} {\bibinfo
  {journal} {Phys. Rev. A}\ }\textbf {\bibinfo {volume} {48}},\ \bibinfo
  {pages} {3837} (\bibinfo {year} {1993})}\BibitemShut {NoStop}%
\bibitem [{\citenamefont {Zuo}\ and\ \citenamefont {Bandrauk}(1995)}]{ZB95}%
  \BibitemOpen
  \bibfield  {author} {\bibinfo {author} {\bibfnamefont {T.}~\bibnamefont
  {Zuo}}\ and\ \bibinfo {author} {\bibfnamefont {A.~D.}\ \bibnamefont
  {Bandrauk}},\ }\href {\doibase 10.1103/PhysRevA.52.R2511} {\bibfield
  {journal} {\bibinfo  {journal} {Phys. Rev. A}\ }\textbf {\bibinfo {volume}
  {52}},\ \bibinfo {pages} {R2511} (\bibinfo {year} {1995})}\BibitemShut
  {NoStop}%
\bibitem [{\citenamefont {Seideman}\ \emph {et~al.}(1995)\citenamefont
  {Seideman}, \citenamefont {Ivanov},\ and\ \citenamefont {Corkum}}]{SIC95}%
  \BibitemOpen
  \bibfield  {author} {\bibinfo {author} {\bibfnamefont {T.}~\bibnamefont
  {Seideman}}, \bibinfo {author} {\bibfnamefont {M.~Y.}\ \bibnamefont
  {Ivanov}}, \ and\ \bibinfo {author} {\bibfnamefont {P.~B.}\ \bibnamefont
  {Corkum}},\ }\href {\doibase 10.1103/PhysRevLett.75.2819} {\bibfield
  {journal} {\bibinfo  {journal} {Phys. Rev. Lett.}\ }\textbf {\bibinfo
  {volume} {75}},\ \bibinfo {pages} {2819} (\bibinfo {year}
  {1995})}\BibitemShut {NoStop}%
\bibitem [{\citenamefont {Chelkowski}\ \emph {et~al.}(1996)\citenamefont
  {Chelkowski}, \citenamefont {Conjusteau}, \citenamefont {Zuo},\ and\
  \citenamefont {Bandrauk}}]{CCZB96}%
  \BibitemOpen
  \bibfield  {author} {\bibinfo {author} {\bibfnamefont {S.}~\bibnamefont
  {Chelkowski}}, \bibinfo {author} {\bibfnamefont {A.}~\bibnamefont
  {Conjusteau}}, \bibinfo {author} {\bibfnamefont {T.}~\bibnamefont {Zuo}}, \
  and\ \bibinfo {author} {\bibfnamefont {A.~D.}\ \bibnamefont {Bandrauk}},\
  }\href {\doibase 10.1103/PhysRevA.54.3235} {\bibfield  {journal} {\bibinfo
  {journal} {Phys. Rev. A}\ }\textbf {\bibinfo {volume} {54}},\ \bibinfo
  {pages} {3235} (\bibinfo {year} {1996})}\BibitemShut {NoStop}%
\bibitem [{\citenamefont {Ben-Itzhak}\ \emph {et~al.}(2008)\citenamefont
  {Ben-Itzhak}, \citenamefont {Wang}, \citenamefont {Sayler}, \citenamefont
  {Carnes}, \citenamefont {Leonard}, \citenamefont {Esry}, \citenamefont
  {Alnaser}, \citenamefont {Ulrich}, \citenamefont {Tong}, \citenamefont
  {Litvinyuk}, \citenamefont {Maharjan}, \citenamefont {Ranitovic},
  \citenamefont {Osipov}, \citenamefont {Ghimire}, \citenamefont {Chang},\ and\
  \citenamefont {Cocke}}]{BWSC08}%
  \BibitemOpen
  \bibfield  {author} {\bibinfo {author} {\bibfnamefont {I.}~\bibnamefont
  {Ben-Itzhak}}, \bibinfo {author} {\bibfnamefont {P.~Q.}\ \bibnamefont
  {Wang}}, \bibinfo {author} {\bibfnamefont {A.~M.}\ \bibnamefont {Sayler}},
  \bibinfo {author} {\bibfnamefont {K.~D.}\ \bibnamefont {Carnes}}, \bibinfo
  {author} {\bibfnamefont {M.}~\bibnamefont {Leonard}}, \bibinfo {author}
  {\bibfnamefont {B.~D.}\ \bibnamefont {Esry}}, \bibinfo {author}
  {\bibfnamefont {A.~S.}\ \bibnamefont {Alnaser}}, \bibinfo {author}
  {\bibfnamefont {B.}~\bibnamefont {Ulrich}}, \bibinfo {author} {\bibfnamefont
  {X.~M.}\ \bibnamefont {Tong}}, \bibinfo {author} {\bibfnamefont {I.~V.}\
  \bibnamefont {Litvinyuk}}, \bibinfo {author} {\bibfnamefont {C.~M.}\
  \bibnamefont {Maharjan}}, \bibinfo {author} {\bibfnamefont {P.}~\bibnamefont
  {Ranitovic}}, \bibinfo {author} {\bibfnamefont {T.}~\bibnamefont {Osipov}},
  \bibinfo {author} {\bibfnamefont {S.}~\bibnamefont {Ghimire}}, \bibinfo
  {author} {\bibfnamefont {Z.}~\bibnamefont {Chang}}, \ and\ \bibinfo {author}
  {\bibfnamefont {C.~L.}\ \bibnamefont {Cocke}},\ }\href {\doibase
  10.1103/PhysRevA.78.063419} {\bibfield  {journal} {\bibinfo  {journal} {Phys.
  Rev. A}\ }\textbf {\bibinfo {volume} {78}},\ \bibinfo {pages} {063419}
  (\bibinfo {year} {2008})}\BibitemShut {NoStop}%
\bibitem [{\citenamefont {Constant}\ \emph {et~al.}(1996)\citenamefont
  {Constant}, \citenamefont {Stapelfeldt},\ and\ \citenamefont
  {Corkum}}]{CSC96}%
  \BibitemOpen
  \bibfield  {author} {\bibinfo {author} {\bibfnamefont {E.}~\bibnamefont
  {Constant}}, \bibinfo {author} {\bibfnamefont {H.}~\bibnamefont
  {Stapelfeldt}}, \ and\ \bibinfo {author} {\bibfnamefont {P.~B.}\ \bibnamefont
  {Corkum}},\ }\href {\doibase 10.1103/PhysRevLett.76.4140} {\bibfield
  {journal} {\bibinfo  {journal} {Phys. Rev. Lett.}\ }\textbf {\bibinfo
  {volume} {76}},\ \bibinfo {pages} {4140} (\bibinfo {year}
  {1996})}\BibitemShut {NoStop}%
\bibitem [{\citenamefont {Wu}\ \emph {et~al.}(2012)\citenamefont {Wu},
  \citenamefont {Meckel}, \citenamefont {Schmidt}, \citenamefont {Kunitski},
  \citenamefont {Voss}, \citenamefont {Sann}, \citenamefont {Kim},
  \citenamefont {Jahnke}, \citenamefont {Czasch},\ and\ \citenamefont
  {D{\"o}rner}}]{WMS12}%
  \BibitemOpen
  \bibfield  {author} {\bibinfo {author} {\bibfnamefont {J.}~\bibnamefont
  {Wu}}, \bibinfo {author} {\bibfnamefont {M.}~\bibnamefont {Meckel}}, \bibinfo
  {author} {\bibfnamefont {L.~P.~H.}\ \bibnamefont {Schmidt}}, \bibinfo
  {author} {\bibfnamefont {M.}~\bibnamefont {Kunitski}}, \bibinfo {author}
  {\bibfnamefont {S.}~\bibnamefont {Voss}}, \bibinfo {author} {\bibfnamefont
  {H.}~\bibnamefont {Sann}}, \bibinfo {author} {\bibfnamefont {H.}~\bibnamefont
  {Kim}}, \bibinfo {author} {\bibfnamefont {T.}~\bibnamefont {Jahnke}},
  \bibinfo {author} {\bibfnamefont {A.}~\bibnamefont {Czasch}}, \ and\ \bibinfo
  {author} {\bibfnamefont {R.}~\bibnamefont {D{\"o}rner}},\ }\href@noop {}
  {\bibfield  {journal} {\bibinfo  {journal} {Nature communications}\ }\textbf
  {\bibinfo {volume} {3}},\ \bibinfo {pages} {1113} (\bibinfo {year}
  {2012})}\BibitemShut {NoStop}%
\bibitem [{\citenamefont {Chelkowski}\ \emph {et~al.}(1998)\citenamefont
  {Chelkowski}, \citenamefont {Foisy},\ and\ \citenamefont {Bandrauk}}]{CFB98}%
  \BibitemOpen
  \bibfield  {author} {\bibinfo {author} {\bibfnamefont {S.}~\bibnamefont
  {Chelkowski}}, \bibinfo {author} {\bibfnamefont {C.}~\bibnamefont {Foisy}}, \
  and\ \bibinfo {author} {\bibfnamefont {A.~D.}\ \bibnamefont {Bandrauk}},\
  }\href {\doibase 10.1103/PhysRevA.57.1176} {\bibfield  {journal} {\bibinfo
  {journal} {Phys. Rev. A}\ }\textbf {\bibinfo {volume} {57}},\ \bibinfo
  {pages} {1176} (\bibinfo {year} {1998})}\BibitemShut {NoStop}%
\bibitem [{\citenamefont {Chelkowski}\ and\ \citenamefont
  {Bandrauk}(1995)}]{CB95}%
  \BibitemOpen
  \bibfield  {author} {\bibinfo {author} {\bibfnamefont {S.}~\bibnamefont
  {Chelkowski}}\ and\ \bibinfo {author} {\bibfnamefont {A.}~\bibnamefont
  {Bandrauk}},\ }\href@noop {} {\bibfield  {journal} {\bibinfo  {journal}
  {Journal of Physics B: Atomic, Molecular and Optical Physics}\ }\textbf
  {\bibinfo {volume} {28}},\ \bibinfo {pages} {L723} (\bibinfo {year}
  {1995})}\BibitemShut {NoStop}%
\bibitem [{\citenamefont {Yu}\ \emph {et~al.}(1998)\citenamefont {Yu},
  \citenamefont {Zuo},\ and\ \citenamefont {Bandrauk}}]{YZB95}%
  \BibitemOpen
  \bibfield  {author} {\bibinfo {author} {\bibfnamefont {H.}~\bibnamefont
  {Yu}}, \bibinfo {author} {\bibfnamefont {T.}~\bibnamefont {Zuo}}, \ and\
  \bibinfo {author} {\bibfnamefont {A.~D.}\ \bibnamefont {Bandrauk}},\
  }\href@noop {} {\bibfield  {journal} {\bibinfo  {journal} {Journal of Physics
  B: Atomic, Molecular and Optical Physics}\ }\textbf {\bibinfo {volume}
  {31}},\ \bibinfo {pages} {1533} (\bibinfo {year} {1998})}\BibitemShut
  {NoStop}%
\bibitem [{\citenamefont {Bandrauk}\ and\ \citenamefont
  {L{\'e}gar{\'e}}(2012)}]{BL12}%
  \BibitemOpen
  \bibfield  {author} {\bibinfo {author} {\bibfnamefont {A.~D.}\ \bibnamefont
  {Bandrauk}}\ and\ \bibinfo {author} {\bibfnamefont {F.}~\bibnamefont
  {L{\'e}gar{\'e}}},\ }in\ \href@noop {} {\emph {\bibinfo {booktitle} {Progress
  in Ultrafast Intense Laser Science VIII}}}\ (\bibinfo  {publisher}
  {Springer},\ \bibinfo {year} {2012})\ pp.\ \bibinfo {pages}
  {29--46}\BibitemShut {NoStop}%
\bibitem [{\citenamefont {Hammes-Schiffer}\ and\ \citenamefont
  {Soudackov}(2008)}]{HS08}%
  \BibitemOpen
  \bibfield  {author} {\bibinfo {author} {\bibfnamefont {S.}~\bibnamefont
  {Hammes-Schiffer}}\ and\ \bibinfo {author} {\bibfnamefont {A.~V.}\
  \bibnamefont {Soudackov}},\ }\href@noop {} {\bibfield  {journal} {\bibinfo
  {journal} {The Journal of Physical Chemistry B}\ }\textbf {\bibinfo {volume}
  {112}},\ \bibinfo {pages} {14108} (\bibinfo {year} {2008})}\BibitemShut
  {NoStop}%
\bibitem [{\citenamefont {Takemoto}\ and\ \citenamefont {Becker}(2010)}]{TB10}%
  \BibitemOpen
  \bibfield  {author} {\bibinfo {author} {\bibfnamefont {N.}~\bibnamefont
  {Takemoto}}\ and\ \bibinfo {author} {\bibfnamefont {A.}~\bibnamefont
  {Becker}},\ }\href {\doibase 10.1103/PhysRevLett.105.203004} {\bibfield
  {journal} {\bibinfo  {journal} {Phys. Rev. Lett.}\ }\textbf {\bibinfo
  {volume} {105}},\ \bibinfo {pages} {203004} (\bibinfo {year}
  {2010})}\BibitemShut {NoStop}%
\bibitem [{\citenamefont {Takemoto}\ and\ \citenamefont {Becker}(2011)}]{TB11}%
  \BibitemOpen
  \bibfield  {author} {\bibinfo {author} {\bibfnamefont {N.}~\bibnamefont
  {Takemoto}}\ and\ \bibinfo {author} {\bibfnamefont {A.}~\bibnamefont
  {Becker}},\ }\href {\doibase 10.1103/PhysRevA.84.023401} {\bibfield
  {journal} {\bibinfo  {journal} {Phys. Rev. A}\ }\textbf {\bibinfo {volume}
  {84}},\ \bibinfo {pages} {023401} (\bibinfo {year} {2011})}\BibitemShut
  {NoStop}%
\bibitem [{\citenamefont {L{\'e}gar{\'e}}\ \emph {et~al.}(2003)\citenamefont
  {L{\'e}gar{\'e}}, \citenamefont {Litvinyuk}, \citenamefont {Dooley},
  \citenamefont {Qu{\'e}r{\'e}}, \citenamefont {Bandrauk}, \citenamefont
  {Villeneuve},\ and\ \citenamefont {Corkum}}]{LLDQ03}%
  \BibitemOpen
  \bibfield  {author} {\bibinfo {author} {\bibfnamefont {F.}~\bibnamefont
  {L{\'e}gar{\'e}}}, \bibinfo {author} {\bibfnamefont {I.}~\bibnamefont
  {Litvinyuk}}, \bibinfo {author} {\bibfnamefont {P.}~\bibnamefont {Dooley}},
  \bibinfo {author} {\bibfnamefont {F.}~\bibnamefont {Qu{\'e}r{\'e}}}, \bibinfo
  {author} {\bibfnamefont {A.}~\bibnamefont {Bandrauk}}, \bibinfo {author}
  {\bibfnamefont {D.}~\bibnamefont {Villeneuve}}, \ and\ \bibinfo {author}
  {\bibfnamefont {P.}~\bibnamefont {Corkum}},\ }\href@noop {} {\bibfield
  {journal} {\bibinfo  {journal} {Phys. Rev. Lett.}\ }\textbf {\bibinfo
  {volume} {91}},\ \bibinfo {pages} {093002} (\bibinfo {year}
  {2003})}\BibitemShut {NoStop}%
\bibitem [{\citenamefont {Abedi}\ \emph {et~al.}(2010)\citenamefont {Abedi},
  \citenamefont {Maitra},\ and\ \citenamefont {Gross}}]{AMG10}%
  \BibitemOpen
  \bibfield  {author} {\bibinfo {author} {\bibfnamefont {A.}~\bibnamefont
  {Abedi}}, \bibinfo {author} {\bibfnamefont {N.~T.}\ \bibnamefont {Maitra}}, \
  and\ \bibinfo {author} {\bibfnamefont {E.~K.~U.}\ \bibnamefont {Gross}},\
  }\href {\doibase 10.1103/PhysRevLett.105.123002} {\bibfield  {journal}
  {\bibinfo  {journal} {Phys. Rev. Lett.}\ }\textbf {\bibinfo {volume} {105}},\
  \bibinfo {pages} {123002} (\bibinfo {year} {2010})}\BibitemShut {NoStop}%
\bibitem [{\citenamefont {Abedi}\ \emph {et~al.}(2012)\citenamefont {Abedi},
  \citenamefont {Maitra},\ and\ \citenamefont {Gross}}]{AMG12}%
  \BibitemOpen
  \bibfield  {author} {\bibinfo {author} {\bibfnamefont {A.}~\bibnamefont
  {Abedi}}, \bibinfo {author} {\bibfnamefont {N.~T.}\ \bibnamefont {Maitra}}, \
  and\ \bibinfo {author} {\bibfnamefont {E.~K.~U.}\ \bibnamefont {Gross}},\
  }\href {\doibase http://dx.doi.org/10.1063/1.4745836} {\bibfield  {journal}
  {\bibinfo  {journal} {The Journal of Chemical Physics}\ }\textbf {\bibinfo
  {volume} {137}},\ \bibinfo {eid} {22A530} (\bibinfo {year}
  {2012})}\BibitemShut {NoStop}%
\bibitem [{\citenamefont {Suzuki}\ \emph {et~al.}(2014)\citenamefont {Suzuki},
  \citenamefont {Abedi}, \citenamefont {Maitra}, \citenamefont {Yamashita},\
  and\ \citenamefont {Gross}}]{SAMYG14}%
  \BibitemOpen
  \bibfield  {author} {\bibinfo {author} {\bibfnamefont {Y.}~\bibnamefont
  {Suzuki}}, \bibinfo {author} {\bibfnamefont {A.}~\bibnamefont {Abedi}},
  \bibinfo {author} {\bibfnamefont {N.~T.}\ \bibnamefont {Maitra}}, \bibinfo
  {author} {\bibfnamefont {K.}~\bibnamefont {Yamashita}}, \ and\ \bibinfo
  {author} {\bibfnamefont {E.~K.~U.}\ \bibnamefont {Gross}},\ }\href {\doibase
  10.1103/PhysRevA.89.040501} {\bibfield  {journal} {\bibinfo  {journal} {Phys.
  Rev. A}\ }\textbf {\bibinfo {volume} {89}},\ \bibinfo {pages} {040501}
  (\bibinfo {year} {2014})}\BibitemShut {NoStop}%
\bibitem [{\citenamefont {Suzuki}\ \emph {et~al.}(2015)\citenamefont {Suzuki},
  \citenamefont {Abedi}, \citenamefont {Maitra},\ and\ \citenamefont
  {Gross}}]{SAMG15}%
  \BibitemOpen
  \bibfield  {author} {\bibinfo {author} {\bibfnamefont {Y.}~\bibnamefont
  {Suzuki}}, \bibinfo {author} {\bibfnamefont {A.}~\bibnamefont {Abedi}},
  \bibinfo {author} {\bibfnamefont {N.~T.}\ \bibnamefont {Maitra}}, \ and\
  \bibinfo {author} {\bibfnamefont {E.}~\bibnamefont {Gross}},\ }\href@noop {}
  {\bibfield  {journal} {\bibinfo  {journal} {arXiv preprint arXiv:1506.04070}\
  } (\bibinfo {year} {2015})}\BibitemShut {NoStop}%
\bibitem [{\citenamefont {Javanainen}\ \emph {et~al.}(1988)\citenamefont
  {Javanainen}, \citenamefont {Eberly},\ and\ \citenamefont {Su}}]{JES88}%
  \BibitemOpen
  \bibfield  {author} {\bibinfo {author} {\bibfnamefont {J.}~\bibnamefont
  {Javanainen}}, \bibinfo {author} {\bibfnamefont {J.~H.}\ \bibnamefont
  {Eberly}}, \ and\ \bibinfo {author} {\bibfnamefont {Q.}~\bibnamefont {Su}},\
  }\href@noop {} {\bibfield  {journal} {\bibinfo  {journal} {Physical Review
  A}\ }\textbf {\bibinfo {volume} {38}},\ \bibinfo {pages} {3430} (\bibinfo
  {year} {1988})}\BibitemShut {NoStop}%
\bibitem [{\citenamefont {Kulander}\ \emph {et~al.}(1996)\citenamefont
  {Kulander}, \citenamefont {Mies},\ and\ \citenamefont {Schafer}}]{KMS96}%
  \BibitemOpen
  \bibfield  {author} {\bibinfo {author} {\bibfnamefont {K.~C.}\ \bibnamefont
  {Kulander}}, \bibinfo {author} {\bibfnamefont {F.~H.}\ \bibnamefont {Mies}},
  \ and\ \bibinfo {author} {\bibfnamefont {K.~J.}\ \bibnamefont {Schafer}},\
  }\href {\doibase 10.1103/PhysRevA.53.2562} {\bibfield  {journal} {\bibinfo
  {journal} {Phys. Rev. A}\ }\textbf {\bibinfo {volume} {53}},\ \bibinfo
  {pages} {2562} (\bibinfo {year} {1996})}\BibitemShut {NoStop}%
\bibitem [{Note1()}]{Note1}%
  \BibitemOpen
  \bibinfo {note} {These probabilities are calculated as defined in~\cite
  {CCZB96}. Here we have chosen $z_I = \pm 15$ for ionization box while the
  dissociation box is defined as $R_D < 10 $.}\BibitemShut {Stop}%
\bibitem [{Note2()}]{Note2}%
  \BibitemOpen
  \bibinfo {note} {For the fully dynamical calculation (labelled TDSE in
  legend) the ionization rate at time $t$ is calculated via $ (\protect
  \qopname \relax o{ln}( p_v(t-T/2)) -\protect \qopname \relax o{ln}(p_v
  (t+T/2)))/T$ and plotted against the corresponding average internuclear
  separation $<R>(t)$ at time $t$. Here $p_v$ is the probability inside the
  ionization box. In case of CN the ionization rate is computed as given in
  \cite {CCZB96}.}\BibitemShut {Stop}%
\bibitem [{Note3()}]{Note3}%
  \BibitemOpen
  \bibinfo {note} {$I(R,t)$ can be rewritten in terms of the exact conditional
  electronic density $|\Phi _R(z,t)|^2=|\Psi |^2/\DOTSI \intop \ilimits@ dz
  |\Psi |^2$ (see ~\cite {AMG10}), i.e, $I(R,t) =\DOTSI \intop \ilimits@ dz
  |\Psi |^2\left [1-\DOTSI \intop \ilimits@ _{z_{I}}dz|\Phi _R(z,t)|^{2}\right
  ]$. Note that $|\Phi _R|^2$ gives the conditional probability of finding
  electrons at $z$ at time $t$, given that the internuclear separation $R$ and
  can be viewed as an exactification of the electronic density for a given
  internuclear separation in quasistatic picture. Therefore, $I(R,t)$ can be
  also viewed as an exact concept analogous to the ionization probability in
  the quasistatic picture.}\BibitemShut {Stop}%
\bibitem [{\citenamefont {Bocharova}\ \emph {et~al.}(2011)\citenamefont
  {Bocharova}, \citenamefont {Karimi}, \citenamefont {Penka}, \citenamefont
  {Brichta}, \citenamefont {Lassonde}, \citenamefont {Fu}, \citenamefont
  {Kieffer}, \citenamefont {Bandrauk}, \citenamefont {Litvinyuk}, \citenamefont
  {Sanderson} \emph {et~al.}}]{BKPB11}%
  \BibitemOpen
  \bibfield  {author} {\bibinfo {author} {\bibfnamefont {I.}~\bibnamefont
  {Bocharova}}, \bibinfo {author} {\bibfnamefont {R.}~\bibnamefont {Karimi}},
  \bibinfo {author} {\bibfnamefont {E.~F.}\ \bibnamefont {Penka}}, \bibinfo
  {author} {\bibfnamefont {J.-P.}\ \bibnamefont {Brichta}}, \bibinfo {author}
  {\bibfnamefont {P.}~\bibnamefont {Lassonde}}, \bibinfo {author}
  {\bibfnamefont {X.}~\bibnamefont {Fu}}, \bibinfo {author} {\bibfnamefont
  {J.-C.}\ \bibnamefont {Kieffer}}, \bibinfo {author} {\bibfnamefont {A.~D.}\
  \bibnamefont {Bandrauk}}, \bibinfo {author} {\bibfnamefont {I.}~\bibnamefont
  {Litvinyuk}}, \bibinfo {author} {\bibfnamefont {J.}~\bibnamefont
  {Sanderson}},  \emph {et~al.},\ }\href@noop {} {\bibfield  {journal}
  {\bibinfo  {journal} {Phys. Rev. Lett.}\ }\textbf {\bibinfo {volume} {107}},\
  \bibinfo {pages} {063201} (\bibinfo {year} {2011})}\BibitemShut {NoStop}%
\end{thebibliography}%

\end{document}